# Enhancing interfacial thermal conductance of amorphous interface by optimized interfacial mass distribution

Lina Yang[1*], Baosheng Yang[1], Baowen Li[2,3*]

[1] School of Aerospace Engineering, Beijing Institute of Technology, Beijing 100081, China

[2] Department of Materials Science and Engineering, Southern University of Science and Technology, Shenzhen, 518055, PR China

[3] Department of Physics, Southern University of Science and Technology, Shenzhen, 518055, PR China

Email: yangln@bit.edu.cn (L. Yang); libw@sustech.edu.cn (B. Li)

# Abstract

Interfacial thermal resistance arises challenges for the thermal management as the modern semiconductors are miniatured to nanoscale. Previous studies found that graded mass distribution in interface can maximumly enhance the interfacial thermal conductance of crystalline interface, however, whether this strategy is effective for amorphous interface is less explored. In this work, graded mass distribution in the amorphous interface between crystalline Si and crystalline Ge is optimized to increase the interfacial thermal conductance by the extended atomistic Green's function method. The results show that atomic mass of 26 amu for one type of atomic mass, and 24 amu and 31 amu for two types of atomic mass in the amorphous interface, can maximumly increase the interfacial thermal conductance. Therefore, the strategy of graded mass distribution is still effective when only considering the atoms in the amorphous interface. In addition, applying the value of the smaller atomic mass of the two sides to the amorphous interface can largely enhance the interface thermal conductance, which is only 0.9% smaller than the maximum value. Further analyses show that atomic mass of 26 amu can increase the phonon modal transmission at high frequency (> 4 THz), and the phonon spectral transmissions are almost the same at low frequency (<2 THz) for different distributions of atomic mass in the amorphous interface in this work. The findings of this work are expected to provide references for optimizing the interfacial thermal conductance of amorphous interfaces for semiconductor devices.

## 1. Introduction

As the modern semiconductor devices are miniatured to nanoscale, interfacial thermal resistance (ITR) arises challenges for the thermal management.[1-5] With sufficient density, solid interfaces play a critical role in determining the overall thermal resistance of nanostructured devices.[6,7] In recent decades, amorphous materials have been widely applied in microelectronics and energy conversion devices such as solar cells[8], transistors[9] and thermoelectric devices[10]. Understanding the thermal transport behavior at amorphous interface is of great importance, however, it has been relatively less explored compared with the extensive study on the thermal management in crystalline interfaces.

The manipulation of interfacial thermal conductance (ITC) of amorphous interface has been studied in previous work. It was found that interfacial coupling strength has strong effect on the ITR at amorphous silicon dioxide/crystalline silicon (a-$SiO_2$/c-Si) interface, with the strong interfacial coupling limit, the ITR reaches a nearly constant value around $0.9\times10^{-9}$ $m^2K/W$ at room temperature.[11] Another work found that the ITR of the c-Si/a-$SiO_2$ $(0.4 \times 10^{-9} \mathrm{m}^2\mathrm{K/W})$ doubles that of the crystalline Si/amorphous silicon (c-Si/a-Si) interface at 500 K.[12] Further, the strain effect shows that the ITR of c-Si/a-$SiO_2$ gradually increases as the strain changes from compression to tension strain[13], moreover, this strain effect is more obvious for interface with weak coupling strength studied by molecular dynamics (MD) simulations. Experimental work on phase change memory devices found that the ITR is substantially changed as $Ge_2Sb_2Te_4$ transitions from cubic to hexagonal crystal structure, resulting in a factor of 4 reduction in the effective thermal conductivity, which is mainly caused by the structural disorder at the interface.[14] In nanocrystalline Si, the ITC of uniform-amorphous $SiO_x$ interlayer is greatly reduced when the interlayer is altered to local crystalline $SiO_x$ by experiment measurements.[15]

Applying graded mass distribution at interface is an effective way to enhance ITC, which has been widely investigated in crystalline interfacial systems. Polanco et al. found that when the atomic mass of the interface is close to the geometric mean of masses of the two contacts, the ITC can reach the maximum value for three-dimensional crystalline interfaces[16]. For the Ar-heavy and Ar crystal system, Zhou et al. compared the ITC of the abrupt interface, mass graded interface and rough interface by nonequilibrium molecular dynamics (NEMD), they found that linearly mass graded interface can increase the ITC by 6-fold compared with the abrupt interface.[17] Similarly, the exponentially mass graded crystalline interface was investigated by Rastgarkafshgarkolaei et al., its ITC initially increases as the number of layers increases but quickly saturates, which is caused by the larger phonon transmission because of the improved DOS bridging.[18]

Interestingly, the 1D atomic model was used to systematically study the enhancement of ITC by graded mass distribution at crystalline interface.[19] Xiong et al. found that the ITC can be increased nearly up to 6-fold when the interface has both geometric graded mass and geometric graded coupling by nonequilibrium Green's function method. Our previous work found the maximum ITC by searching the optimum mass distribution of the interface by combining the atomistic Green's function (AGF) and machine learning method.[20] The optimal mass distribution is also graded, but different from linear and geometric graded distribution. Though the strategy of graded mass distribution has been broadly used in crystalline system, few work applied it to manipulate the ITC of amorphous interface (a-M).

In this work, the ITC of the c-Si/a-M/c-Ge system with graded atomic mass distribution in the amorphous interface is investigated by extended atomistic Green's function method. One type and two types of atomic mass are applied in the amorphous interface to maximumly increase the ITC, and the corresponding optimized atomic mass distribution are compared with other atomic mass distribution. To confirm the results, the interfacial thermal conductances of the c-Si/a-M/c-$Si^{14}$ system are additionally investigated, where c-$Si^{14}$ is the Si isotope with atomic mass of 14 amu. Finally, the phonon modal transmission and spectral phonon transmission are analyzed to understand the underlying physics.

## 2. Model and Method

To study the thermal transport of amorphous interface, the system of c-Si/a-M/c-Ge is built as shown in Figure 1 (a). The atomic masses of Si and Ge atoms are 27.89 and 72.61 amu, respectively. The atomic mass of the atoms in a-M can be varied to manipulate the interfacial thermal conductance. The length of the amorphous interface (yellow atoms) is 43.4 Å, the cross section of the system is 16.3 Å×16.3 Å. Here, the interaction between Si and Ge atoms is assumed to be identical, which has less significant effect than the mass mismatch.[21,22] Tersoff potential[23] is used to describe the interaction between atoms in the system. Periodic boundary conditions are applied in *y* and *z* directions. The heat flux is in *x* direction.

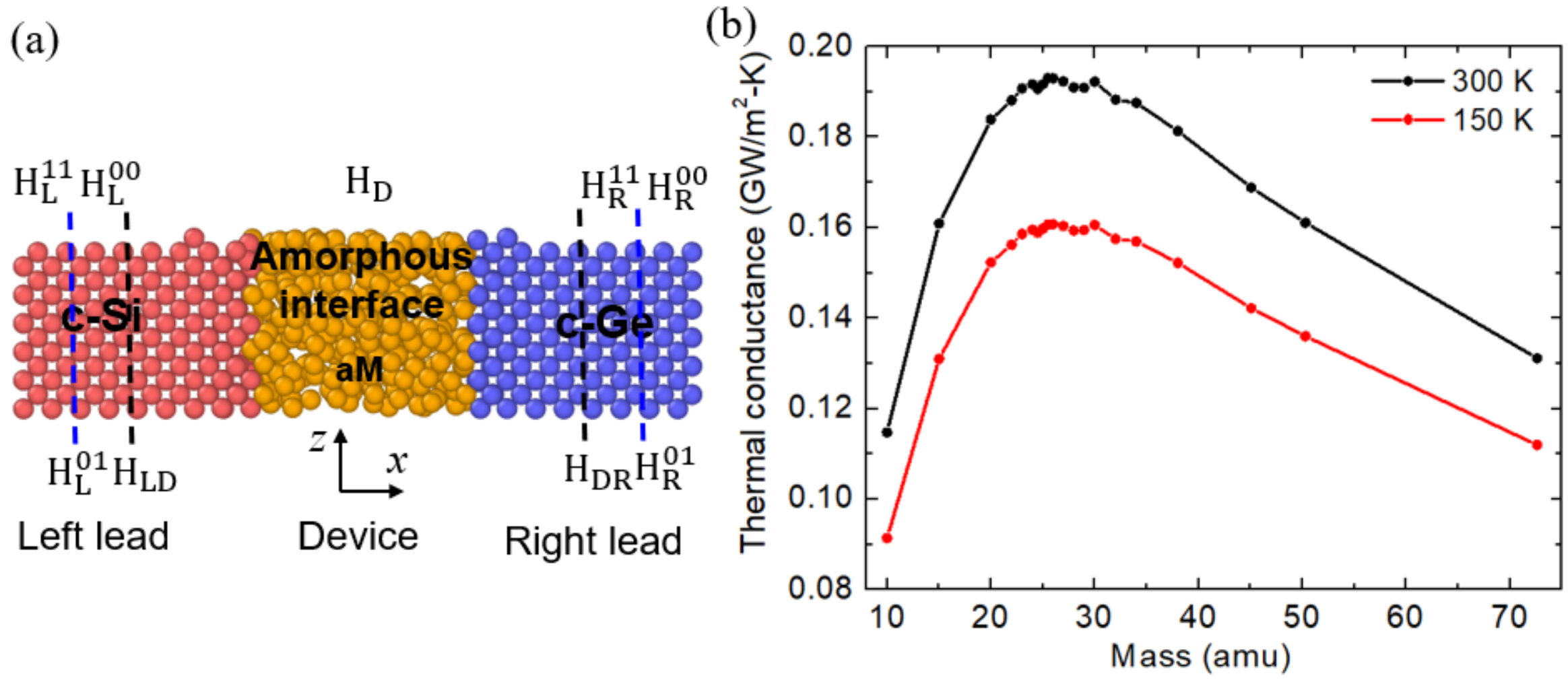

Figure 1. (a) The structure of amorphous interface in the c-Si/a-M/c-Ge system. The pink atoms are c-Si, the blue atoms are c-Ge, and the atoms in the amorphous interface are in yellow color. The atomic mass of the atoms in amorphous interface can be varied. The length of the amorphous interface is 43.4 Å. The system is divided into three parts by the black dashed line: left lead, right lead and device. (b) The interfacial thermal conductance of the amorphous interface with one type of atomic mass (aM). The value of the atomic mass in amorphous interface is changed from 10 to 72.61 amu.

The amorphous interface is created by the melt quenching method[24,25], which is performed by LAMMPS (large-scale atomic/molecular massively parallel simulator)[26]. The time step of MD is 0.5 fs. The procedure follows the steps in Ref.[27]. Initially, the atoms in the interface are melted at 3600 K by Nosé-Hoover thermostat for 0.5 ns, then the interface is quickly quenched to 1000 K at the rate of 860 K/ps. Finally, the interface is annealed at 1000 K for 0.5 ns, then quickly quenched to 20 K at the rate of 160 K/ps. To get the position of atoms in the amorphous interface, an equilibration is first performed for 2.5 ns in an NVT ensemble, then the atom positions are averaged over the next 2.5 ns.

The phonon modal transmission coefficients are calculated by the mode-resolved AGF

formalism[21,28] which is developed in based on the AGF method[29,30]. As shown in Figure 1 (a), the system is divided into three interacting parts by the black dashed line: the left and right semi-infinite leads and the device. The equations of motion of this system can be written by the following matrix form:

$$(\omega^2 \boldsymbol{I} - \boldsymbol{H})\boldsymbol{\psi} = 0 \tag{1}$$

where $\omega$ is the radial frequency, **I** is the identity matrix, $\boldsymbol{\Psi}$ are the eigenvectors of the total system, and **H** is the harmonic matric representing the atomic interactions,

$$\boldsymbol{H} = \begin{bmatrix} \ddots & \vdots & \vdots & & & & \\ \cdots & \mathbf{H}_{\mathbf{L}}^{\mathbf{11}} & \mathbf{H}_{\mathbf{L}}^{\mathbf{00}} & & & & \\ \cdots & \mathbf{H}_{\mathbf{L}}^{\mathbf{10}} & \mathbf{H}_{\mathbf{L}}^{\mathbf{00}} & \boldsymbol{H}_{LD} & 0 & & \\ & 0 & \boldsymbol{H}_{DL} & \boldsymbol{H}_{D} & \boldsymbol{H}_{DR} & 0 & \\ & & 0 & \boldsymbol{H}_{RD} & \mathbf{H}_{\mathbf{R}}^{\mathbf{00}} & \mathbf{H}_{\mathbf{R}}^{\mathbf{01}} & \cdots \\ & & & & \mathbf{H}_{\mathbf{R}}^{\mathbf{10}} & \mathbf{H}_{\mathbf{R}}^{\mathbf{11}} & \cdots \\ & & & & \vdots & \vdots & \ddots \end{bmatrix}$$

where $\boldsymbol{H}_D$ is the dynamical matrix of the amorphous interface, $\boldsymbol{H}_{LD}$ ($\boldsymbol{H}_{RD}$) is the coupling matrix between the left (right) contact and the interface. Because of translational invariance in $y$ and $z$ directions, the Fourier transform is applied with 28×28 wave-vector grids. The phonon spectral transmission $\Xi(\boldsymbol{\omega})$ is calculated by solving the device retarded Green's function. The mode-resolved transmission matrix $\boldsymbol{t}$ is computed through constructing the Bloch matrices. The phonon modal transmission, $\Xi_i(\boldsymbol{\omega}) = \sum_j |\boldsymbol{t}_{ij}|^2$, is for phonon mode $\boldsymbol{i}$ transmitting through the interface. The phonon spectral transmission $\Xi(\boldsymbol{\omega}) = \sum_i \Xi_i(\boldsymbol{\omega})$. The thermal conductance per unit area $\sigma(T)$ for the device region is calculated by the Landauer formalism: $\sigma(T) = \frac{1}{2\pi A}\int_0^{\omega_{max}} \hbar\omega \frac{\partial n}{\partial T}(\omega, T)\Xi(\omega)d\omega$. A is the transverse area, $n(\omega, T)$ is the Bose-Einstein distribution, T is the temperature.

## 3. Results

Firstly, we study the system with one type of atomic mass in the amorphous interface which are the atoms in yellow color in Figure 1 (a). The value of atomic mass (noted as aM) is changed from 10 to 72.61 amu. The ITC of the system with one type of atomic mass in amorphous interface is shown in Figure 1 (b). The atomic mass corresponding to maximum ITC ($ITC_{max}$), noted as $aM_{max}$, is 26 and 25.5 amu for temperature at 150 K and 300 K, respectively, which is smaller than both the atomic masses of the two leads (Si and Ge) and does not follow the arithmetic mean[17] or the geometric mean[18] of atomic masses of the two leads. To further confirm these results, the ITC of the c-Si/a-M/c-Ge system with one type of atomic mass in amorphous interface is studied by NEMD method, which is shown in Figure R1 in the Supplementary Materials. Additionally, the ITC is very close when aM is in the range of 23 to 32. When aM is set as the value of Si (27.98 amu), the ITC is only 0.9% smaller than the $ITC_{max}$. Therefore, when the atoms in the amorphous interface has the smaller atomic mass of the two leads, the ITC of the c-Si/a-M/c-Ge system can be largely enhanced.

To further increase the ITC of the c-Si/a-M/c-Ge system, two types of atomic mass are set in the amorphous interface, the structure is shown in Figure 2 (a). The atomic mass of the atoms in the amorphous interface close to Si is noted as aM1 (pink atoms), and close to Ge is noted as aM2 (blue atoms). The value of aM1 is from 23.5 to 26.5 amu, while the value of aM2 is from 26 to 32.5 amu, the increase step is 0.5 amu, which are shown in Figure 2 (b). A Mass ID is given to a set of aM1 and aM2, totally 98 sets of aM1 and aM2 are studied.

The ITC of the c-Si/a-M/c-Ge system with two types of atomic mass versus Mass ID is shown in Figure 2 (c) and (d) for temperatures at 150 K and 300 K, respectively. The results show that aM1=24 (noted as $aM1_{max}$) and aM2=31 (noted as $aM2_{max}$), corresponding Mass ID=25, can result in maximum ITC, which is neither the linear distribution (corresponding aM1=42.85 amu and aM2=57.73 amu) nor the exponential distribution (corresponding aM1=36.72 amu and aM2=50.59 amu) of the atomic masses of the two leads. However, $aM1_{max}$ is smaller than $aM2_{max}$, and $aM_{max}$ is between $aM1_{max}$ and $aM2_{max}$, which still follows the graded mass distribution form when only considering the atoms in the amorphous interface.

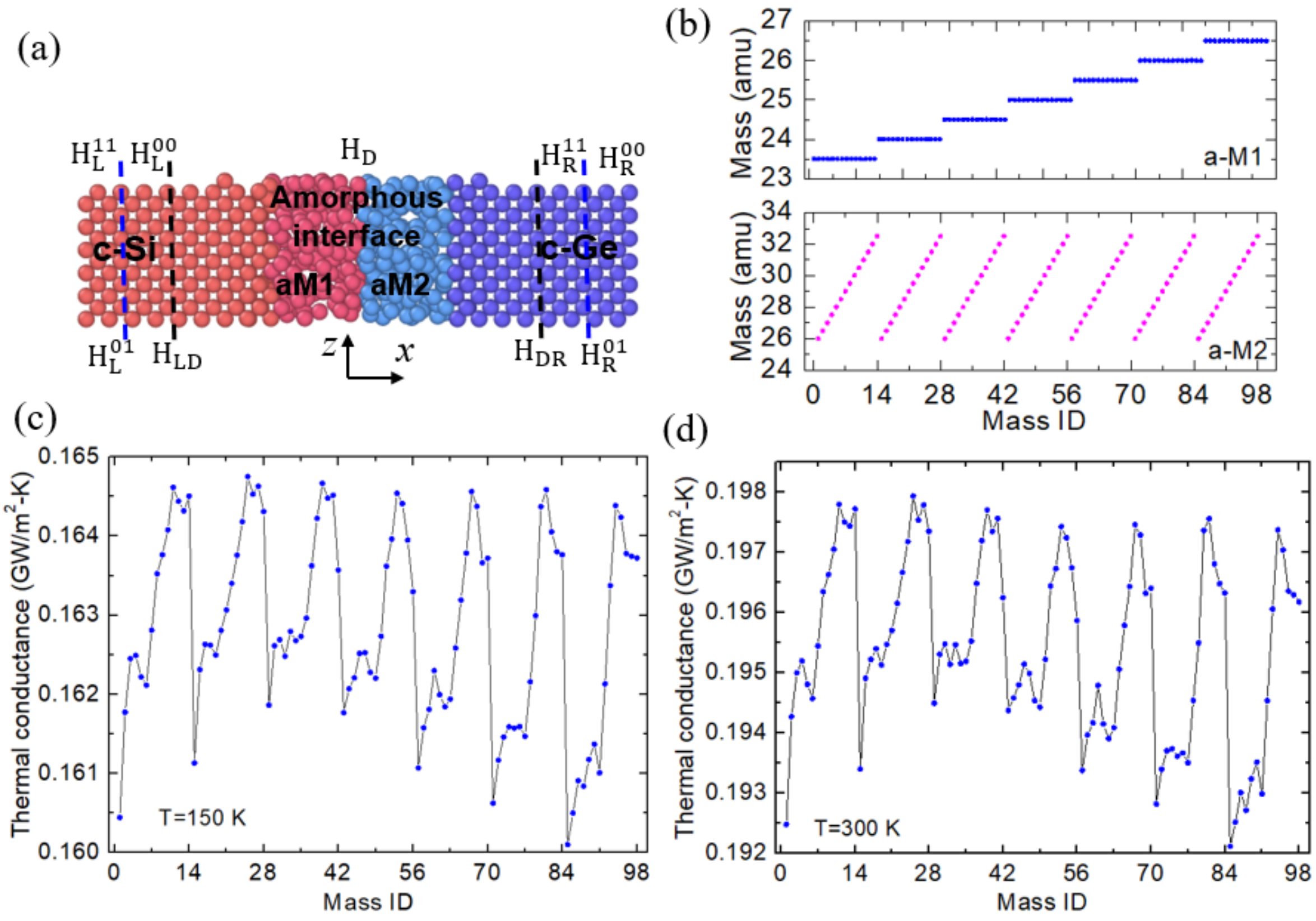

Figure 2 (a) The structure of the c-Si/a-M/c-Ge system with two types of atomic masses (aM1 and aM2) in the amorphous interface. (b) The value of aM1 and aM2. A set of aM1 and aM2 is given a Mass ID, totally 98 sets of aM1 and aM2 are studied. The interfacial thermal conductance versus Mass ID for temperatures at 150 K (c) and 300 K (d).

To compare the enhancement of ITC by different atomic mass distribution in amorphous interface, the ITC of the c-Si/a-M/c-Ge system with one type of atomic mass (Figure 3(a)) and two types of atomic mass (Figure 3 (b)) for temperature at 150 K are investigated. To show the effect of amorphous interface on the reduction of ITC, the ITC of crystalline interface of the c-Si/c-Ge system (blue dashed line in Figure 3) is calculated based on the phonon transmission in Ref.[20]. For one type of atomic mass in amorphous interface, the aM is set to be $aM_{max}$ (26 amu), aSi (atomic mass of Si, 27.89 amu), Exp (geometric mean of the atomic masses of the two leads, 45.07 amu), Linear (arithmetic mean of the atomic masses of the two leads, 50.29 amu), and aGe (atomic mass of Ge, 72.61 amu). The ITC corresponding to $aM_{max}$ and aSi are pretty close, while the Exp has better performance than Linear. The ITC corresponding to aGe is the smallest. For two types of atomic masses in amorphous interface, the aM1 and aM2 are set to be $aM1_{max}$/ $aM2_{max}$ (24 amu/31 amu), $Max_{ref}$ (according to the optimized distribution in Ref.[20], 34.67 amu/52.52 amu), Exp(exponential distribution, 36.71 amu/50.59 amu), Linear (linear distribution, 42.85 amu/57.73 amu), and aSi/aGe (atomic mass of Si and Ge, 27.89 amu/72.61 amu). Interestingly, the ITC is only slightly increased (2.5%) by $aM1_{max}$/ $aM2_{max}$ compared with by $aM_{max}$. For comparison, the ITC of different types of systems including crystalline systems, crystalline interfaces and amorphous interfaces are shown in Figure R2 in the Supplementary Material. Therefore, two types of atomic mass do not show obvious advantages in the enhancement of ITC of amorphous interface compared with one type of atomic mass.

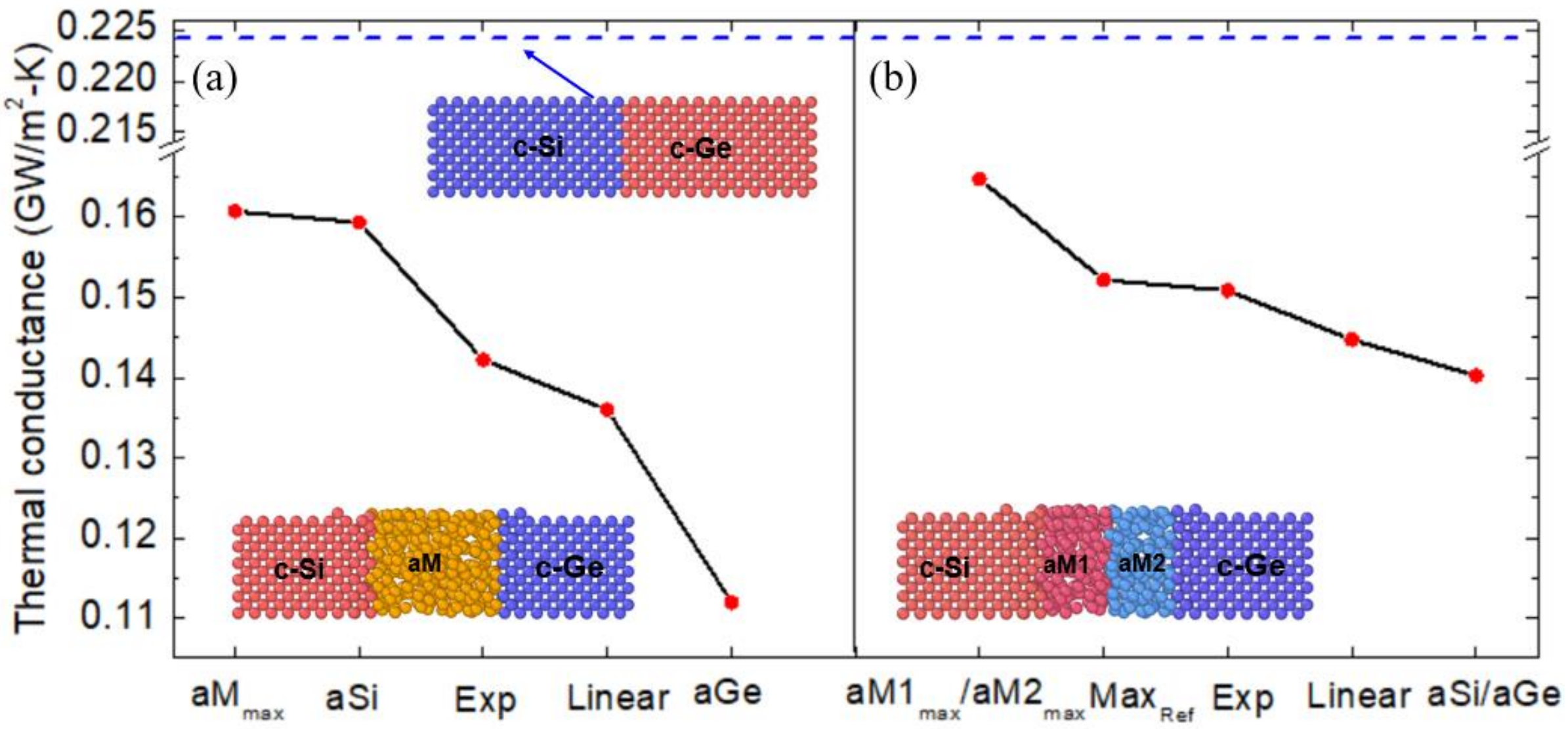

Figure 3 (a) Interfacial thermal conductance of the c-Si/a-M/c-Ge system with one type of mass for temperature at 150 K. The aM is set to be $aM_{max}$ (26 amu), aSi (atomic mass of Si, 27.89 amu), Exp (geometric mean of the atomic mass of the two leads, 45.07 amu), Linear (arithmetic mean of the atomic mass of the two leads, 50.29 amu), and aGe (atomic mass of Ge, 72.61 amu). (b) Interfacial thermal conductance of the c-Si/a-M/c-Ge system with two types of atomic mass for the temperature at 150 K. The aM1 and aM2 are set to be $aM1_{max}$/ $aM2_{max}$ (24 amu/31 amu), $Max_{ref}$ (according to the optimized distribution in Ref.[20], 34.67 amu/52.52 amu), Exp(exponential distribution, 36.71 amu/50.59 amu), Linear (linear distribution, 42.85 amu/57.73 amu), and aSi/aGe (atomic mass of Si and Ge, 27.89 amu/72.61 amu). The ITC of crystalline interface of the c-Si/c-Ge system at 150 K (blue dashed line) is shown for comparison.

To further understand the enhancement of ITC, the phonon spectral transmission and phonon modal transmission versus frequency are compared in Figure 4(a), 4(b) and 4(c), respectively. For low frequency phonons (<2.0 THz), the spectral transmissions are almost the same for amorphous interface with atomic masses of $aM_{max}$, aSi, aGe, $aM1_{max}$/ $aM2_{max}$ and aSi/Ge which have the same meaning as that in Figure 3. It also shows that when phonon frequency is less than 4 THz, the spectral transmissions are close for atomic mass of $aM_{max,}$ aSi, and $aM1_{max}$/ $aM2_{max}$. Compared with atomic mass of aSi, $aM1_{max}$/ $aM2_{max}$ slightly increases the spectral transmission around frequencies at 4.5 THz, 6 THz and 8.5 THz, which leads to the small increase of ITC. Further, phonon modal transmissions are analyzed. As shown in Figure 4 (b) and (c), though atomic masses of aGe and aSi/aGe cause a small reduction of modal transmission at high frequency, spectral transmission is largely reduced (shown in Figure 4 (a)), which leads to the reduction of ITC.

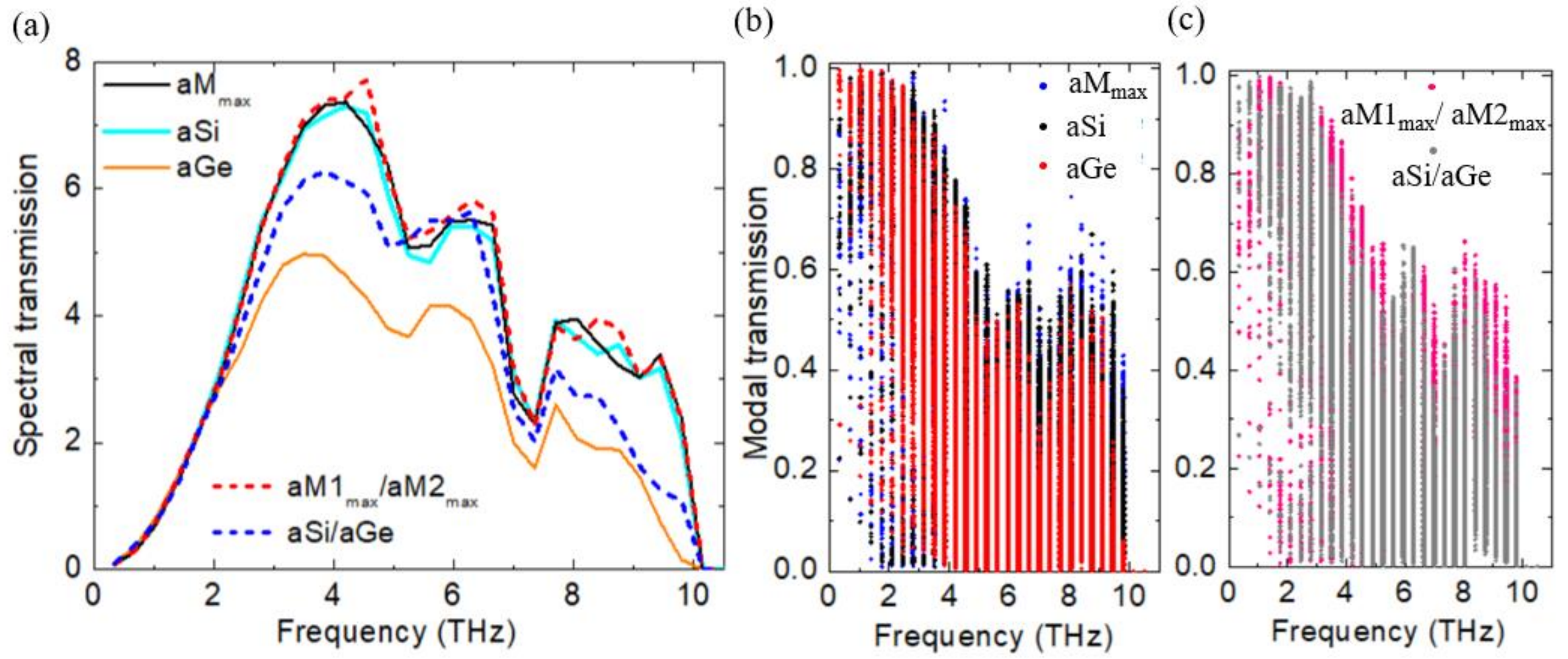

Figure 4 (a) Phonon spectral transmission across the amorphous interface in the c-Si/a-M/c-Ge system. Phonon modal transmission across the amorphous interface with atomic mass of $aM_{max}$, aSi, aGe (b) and $aM1_{max}$/ $aM2_{max}$, aSi/Ge (c) in the c-Si/a-M/c-Ge system. The meanings of $aM_{max}$, aSi, aGe, $aM1_{max}$/ $aM2_{max,}$ and aSi/Ge are the same as that in Figure 3.

For the c-Si/a-M/c-Ge system, atomic mass of Si in the amorphous interface can largely increase the ITC in Figure 1(b), therefore, the smaller atomic mass of the two leads can be used in the amorphous interface to greatly enhance the ITC. To verify if this finding can be used in other similar systems, the c-Si/a-M/c-Si[14] system in Figure 5 (a) is additionally investigated, where c-$Si^{14}$ is the Si isotope with atomic mass of 14 amu. Following the settings in Figure 1(a), one type of atomic mass in the amorphous interface is used in the c-Si/a-M/c-$Si^{14}$ system, with values varying from 5 amu to 30 amu. The ITC for the c-Si/a-M/c-$Si^{14}$ system is shown in Figure 5 (a). The $aM_{max}$ corresponding to $ITC_{max}$ is 13 amu for temperature at both 150 K and 300 K. The value of ITC is close when atomic mass is between 12 amu to 16 amu, so atomic mass of $Si^{14}$ (note as $aSi^{14}$) can largely increase the ITC. In addition, the spectral transmissions across the amorphous interface with atomic mass of $aM_{max}$ (13 amu), aSi and $aSi^{14}$ are close when phonon frequency is less than 3.0 THz, which is shown in Figure 5(b). $aM_{max}$ and $aSi^{14}$ cause larger spectral transmission at high frequency (>3THz) compared with aSi, which leads to the enhancement of ITC. For comparison, the spectral transmission of crystalline interface, c-Si/c-Ge and c-Si/c-$Si^{14}$, are also shown in Figure 5 (b). Therefore, the results of the c-Si/a-M/c-$Si^{14}$ system further confirm that the smaller atomic mass of the two leads can greatly enhance the ITC of the amorphous interface.

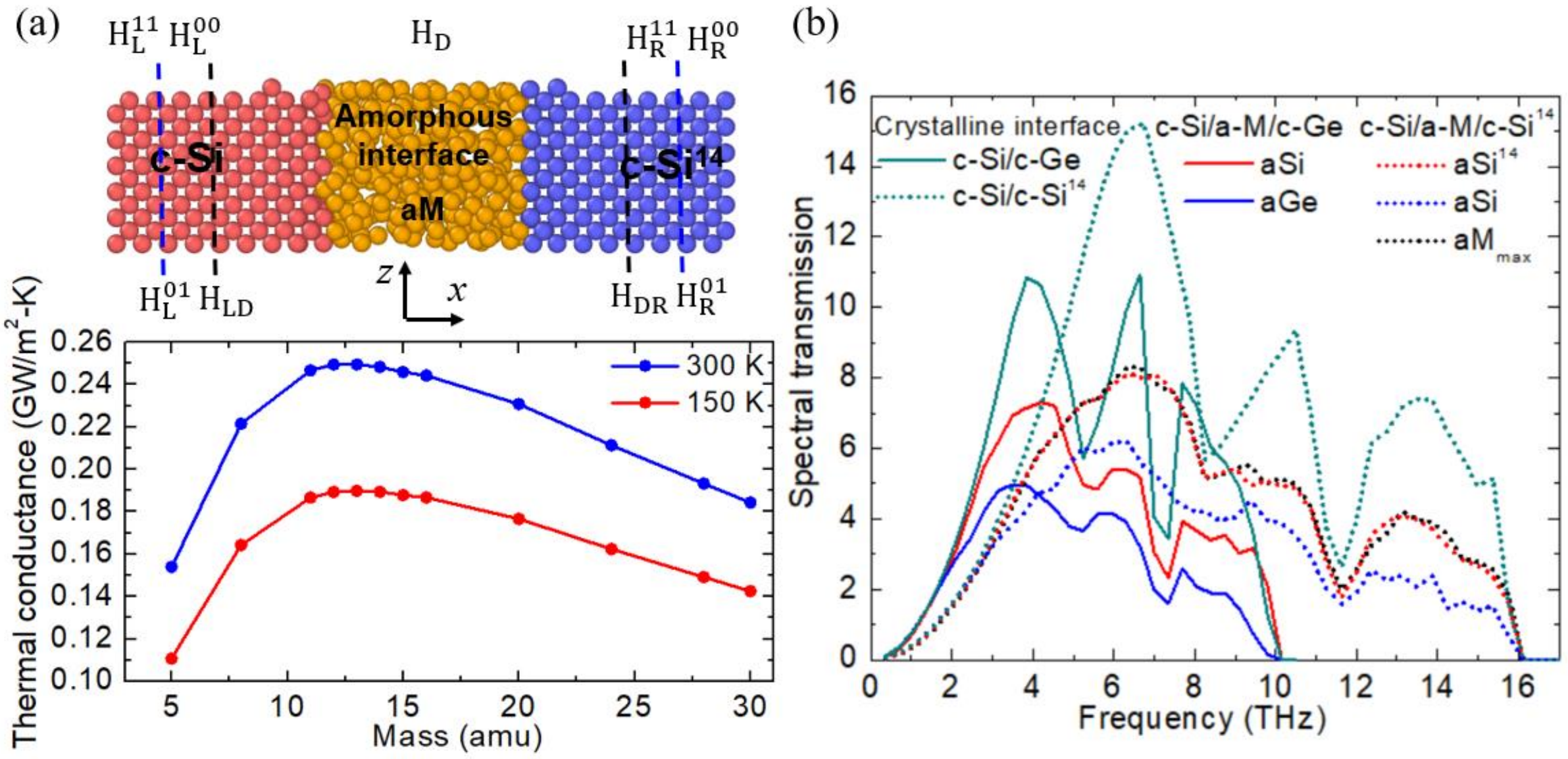

Figure 5. (a) The structure of amorphous interface in the c-Si/a-M/c-$Si^{14}$ system and the ITC of the amorphous interface with one type of atomic mass. The blue atoms are c-$Si^{14}$ which is the Si isotope with atomic mass of 14 amu, all the settings are the same as that in Figure 1(a). The value of the atomic mass in amorphous interface is changed from 5 to 30 amu. (b) The phonon spectral transmission of the c-Si/a-M/c-$Si^{14}$ system. The spectral transmission of crystalline interfaces (c-Si/c-Ge and c-Si/c-$Si^{14}$) and amorphous interfaces (c-Si/a-M/c-Ge) are also shown for comparison.

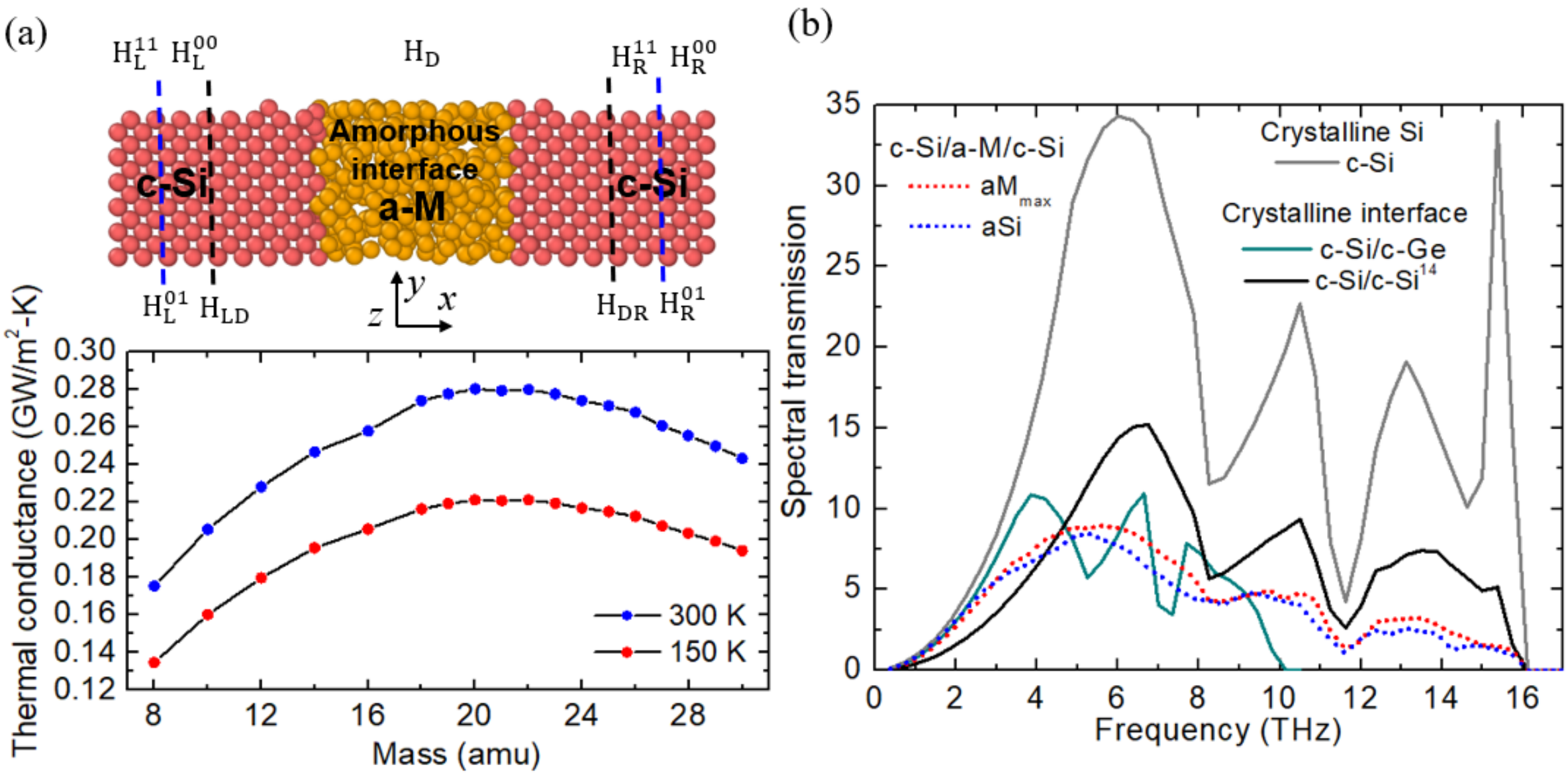

Figure 6. (a) The structure of the amorphous interface in the c-Si/a-M/c-Si system and the ITC of the amorphous interface in the c-Si/a-M/c-Si system with one type of atomic mass. All the settings are the same as that in Figure 1(a). The value of the atomic mass in the amorphous interface is changed from 8 to 30 amu. (b) The phonon spectral transmission for the amorphous interface in the c-Si/a-M/c-Si system. The spectral transmission of the crystalline interface (c-Si/c-Ge and c-Si/c-Si[14]) and crystalline Si are also shown for comparison.

For most previous studies, the two leads have different atomic masses, for example, the Si/Ge system and Ar/ Ar-heavy system, manipulating the atomic mass in the interface can increase their ITC. [16] [17] [18] Whether this strategy is effective for two leads with the same atomic mass has not been investigated. For further investigation, the c-Si/a-M/c-Si system (Figure 6(a)) is studied, and all the settings are the same as that in Figure 1(a). The value of the atomic mass in the amorphous interface is changed from 8 to 30 amu, the corresponding ITC presented in Figure 6 (a) firstly increases, then decreases after reaching the $ITC_{max}$. The $aM_{max}$ is 20 amu for temperature at both 150 K and 300 K, which can maximumly increase the ITC. However, the $aM_{max}$ (20 amu) is smaller than the atomic mass of the two leads. The ITC corresponding to atomic mass of Si (27.89 amu) is 6% smaller than the $ITC_{max}$ for temperature at 150 K. Further analyses show that the $aM_{max}$ can increase the phonon spectral transmission for phonon with larger frequency (>3 THz) as shown in Figure 6(b), which accounts for the increase of ITC. The spectral transmission of the crystalline interface (c-Si/c-Ge and c-Si/c-Si[14]) and crystalline Si are also shown for comparison. Therefore, manipulating the atomic mass in the amorphous interface in the c-Si/a-M/c-Si system can increase the ITC.

## 4. Conclusions

In this work, the interfacial thermal conductance of amorphous interface in the c-Si/a-M/c-Ge system is studied by applying atomistic Green's function method. The results show that the $aM_{max}$ maximumly enhancing the ITC is 26 amu when only one type of atomic mass is used in the amorphous interface, which is smaller than both the atomic masses of the two leads (Si and Ge). When two types of atomic mass are used in the amorphous interface, the $aM1_{max}$ and $aM2_{ma}$ are 24 amu and 31 amu for getting the maximum ITC, which is neither the linear distribution nor the exponential distribution of the atomic masses of the two leads. Interestingly, $aM1_{max}$ is smaller than $aM2_{max}$, and $aM_{max}$ is between $aM1_{max}$ and $aM2_{max}$, therefore, it still shows a graded mass distribution form when only considering the atoms in amorphous interface. Especially, the $ITC_{max}$ corresponding to $aM1_{max}/aM2_{max}$ is only 2.4% larger than that corresponding to $aM_{max}$ for temperature at 150 K. Additionally, when the atomic mass in the amorphous interface is set as the value of the smaller atomic mass of the two leads, the ITC is only 0.9% smaller than the $ITC_{max.}$for the c-Si/a-M/c-Ge system. The analyses of the phonon modal transmission by the extended atomistic Green's function method show that the atomic mass of $aM_{max}$ and $aM1_{max}$/ $aM2_{max}$ slightly increases the phonon modal transmission at high frequency (> 4 THz) compared with atomic mass of aSi, which leads to the small increase of ITC. When frequency is less than 2 THz, the spectral transmissions are almost the same as for all the settings of the atomic mass in the amorphous interface in this study.

## 5. Acknowledgements

This work is sponsored by the National Natural Science Foundation of China (Grant No.12004033) (L.Y.).